\documentstyle[seceq,epsf,mbf,wrapft]{ptptex}

\def\gsim{\,$\raise0.3ex\hbox{$>$}\llap{\lower0.8ex\hbox{$\sim$}}$\,}
\def\lsim{\,$\raise0.3ex\hbox{$<$}\llap{\lower0.8ex\hbox{$\sim$}}$\,}

\title{
Ground-State Phase Diagram of Frustrated Anisotropic Quantum Spin Chains
}

\author{
Toshiya {\sc Hikihara}$^{1,}$\footnote{E-mail: HIKIHARA.Toshiya@nims.go.jp},
Makoto {\sc Kaburagi}$^2$ and Hikaru {\sc Kawamura}$^3$
}

\inst{
$^1$National Institute for Materials Science, Sengen 1-2-1, Tsukuba 305-0047
\\
$^2$Faculty of Cross-Cultural Studies, Kobe University, Tsurukabuto, Nada,
Kobe 657-8501
\\
$^3$Department of Earth and Space Science, Graduate School of Science, 
Osaka University, Toyonaka, Osaka 560-0043
}

\recdate{
}

\abst{
Recent studies on the frustrated quantum spin chains 
with easy-plane anisotropy are reviewed.
We are particularly interested in novel ^^ ^^ chiral" phases 
characterized by the spontaneous breaking of the parity symmetry.
The ground-state phase diagrams of the chains are discussed.
}

\begin{document}

\maketitle

\section{Introduction}

The low-energy properties of one-dimensional quantum spin systems 
with frustrating interactions have been studied extensively for many years. 
This is mainly because they exhibit a variety of 
ground-state phases accompanied with various types of 
spontaneous symmetry breaking, e.g., the dimer phase 
with the translational symmetry breaking.
More recently novel phases have been found in the ground state 
of frustrated quantum spin chains with easy-plane anisotropy:
\cite{Ner,Leche,Kole,KKH,HKK1,HKK2,HKK3,Nishi} 
The chains exhibit ^^ ^^ chiral" ordered phases, 
characterized by the nonzero value of the $z$-component of 
the total vector chirality and the absence of 
the helical spin long-range order (LRO),
\begin{eqnarray}
O_\kappa &=& \frac{1}{LS^2} \sum_l \kappa_l \ne 0 
~~~~~~~ (\kappa_l = S^x_lS^y_{l+1}-S^y_lS^x_{l+1} 
  = \left[ \mib{S}_l \times \mib{S}_{l+1} \right]_z), \label{eq:Ochl} \\
\mib{m}(q) &=& \frac{1}{LS} \sum_l \mib{S}_l \exp(iql) = 0
~~~~~~~({\rm for}~\forall q), \label{eq:hel}
\end{eqnarray}
where $\mib{S}_l$ is spin-$S$ operator at $l$-th site and 
$L$ is the total number of spins.
In the chiral ordered phases, the parity symmetry is broken spontaneously 
while the time-reversal and translational symmetries are preserved.
From analytic\cite{Ner,Leche,Kole} and 
numerical\cite{KKH,HKK1,HKK2,HKK3,Nishi} studies,
it has been revealed that the chiral ordered phases appear in a broad region 
of the ground-state phase diagram.

The purpose of this paper is to review recent progress in 
the understanding of the ground-state properties of 
the frustrated anisotropic quantum spin chains, 
especially focusing on the possible chiral ordered phases.
Results on the ground-state phase diagram of the frustrated $XXZ$ chain 
for general spin-size $S$ are summarized in the next section (\S 2).
We also discuss how the phase diagram changes as $S$ increases.
The phase diagram of the $S = 1$ frustrated Heisenberg chain with 
the single-ion-type anisotropy are discussed in \S 3.
The final section (\S 4) is devoted to a summary and discussion.

\section{Phase Diagram of Frustrated Spin-$S$ $XXZ$ Chain}
In this section, we consider the ground-state phase diagram of 
the frustrated spin-$S$ $XXZ$ chain.
The Hamiltonian of the model has the form,
\begin{equation}
{\cal H} = \sum_{\rho=1,2} J_\rho \sum_l 
\left( S^x_l S^x_{l+\rho} + S^y_l S^y_{l+\rho} 
+ \Delta S^z_l S^z_{l+\rho} \right),
\label{eq:HamXXZ}
\end{equation}
where $\Delta$ is the anisotropy parameter of the exchange couplings.
We consider the case of antiferromagnetic couplings $J_1 > 0$ and $J_2 > 0$ 
and easy-plane anisotropy $0 \le \Delta < 1$.
In the classical limit $S \to \infty$, the ground-state phase diagram 
of the chain has been well understood.
The chain exhibits the magnetic LRO $\mib{m}(q) \ne 0$ for arbitrary 
$j \equiv J_2/J_1$.
The LRO is of the N\'{e}el type $q = \pi$ for $j < 1/4$ 
while it becomes of helical type for $j > 1/4$ with a incommensurate 
wave number $q = \cos^{-1}(-1/4j)$.
In the helical phase, the LRO of the vector chirality $O_{\kappa} \ne 0$ 
is also present so that both the time-reversal and parity symmetries 
are broken spontaneously.

In the quantum case where $S$ is finite, on the other hand, 
it seems to be well established that no magnetic LRO appears 
at least for $0 \le \Delta \le 1$ because of quantum fluctuations.
We note that the absence of the magnetic LRO in the spin chain 
(\ref{eq:HamXXZ}) has been proven rigorously in the $XY$ ($\Delta = 0$) 
and Heisenberg ($\Delta = 1$) cases.\cite{Momoi}
By contrast, no theorem prohibiting the LRO of the vector chirality 
$O_\kappa \ne 0$ has been known.
Thus, in principle, there remains a possibility that the chiral ordered phase 
appears in the frustrated quantum chain (\ref{eq:HamXXZ}).
Furthermore, it has been known that the chain exhibits 
the so-called spin-fluid (SF) phase 
and phases with topological order, i.e., 
the dimer phase\cite{dim} for half-odd-integer $S$, 
and the Haldane\cite{Hal} and 
double-Haldane (DH)\cite{DH} phases for integer $S$.

In order to search for the possible chiral ordered phase,
many studies have been performed.
Using the bosonization technique combined with a mean-field analysis, 
Nersesyan {\it et al.} have pointed out that the $S = 1/2$ chain 
(\ref{eq:HamXXZ}) in the case of $\Delta = 0$ and $j \gg 1$ 
exhibits the chiral ordered phase with gapless excitations.
This conclusion has been generalized later for the case of $S \ge 1$
by extending the bosonization method.\cite{Leche}
The schematic phase diagram for general integer $S$ 
including the chiral phases has been proposed 
by means of a large-$S$ approach.\cite{Kole}

In parallel with these analytic approaches, the possibility of 
the chiral ordered phase has been pursued in numerical studies 
by the present authors.\cite{KKH,HKK1,HKK2,HKK3}
Since the incommensurate character of the system for large $j$ 
causes a severe finite-size effect,\cite{Ali}
we have employed the density-matrix renormalization-group (DMRG)
method,\cite{White} which can treat extremely large systems 
and obtain results in the thermodynamic limit directly.
By calculating spin, chiral, and topological correlation functions 
associated with each order parameter 
and examining their long-distance behaviors, 
we have determined the ground-state phase diagram of the frustrated chain 
for $S = 1/2, 1, 3/2$, and $2$.
Then, we have found that there exist different types of the chiral phase 
with gapless and gapful excitations.

From these intensive studies, we now consider that 
the appearance of the chiral ordered phase in the frustrated quantum $XXZ$ 
chain has been established.
Details of the results for integer and half-odd-integer $S$ 
and the $S$-dependence of the phase diagram 
are summarized in the following subsections.

\begin{wrapfigure}{r}{6.6cm}   
  \epsfxsize=6.6cm
  \epsfbox{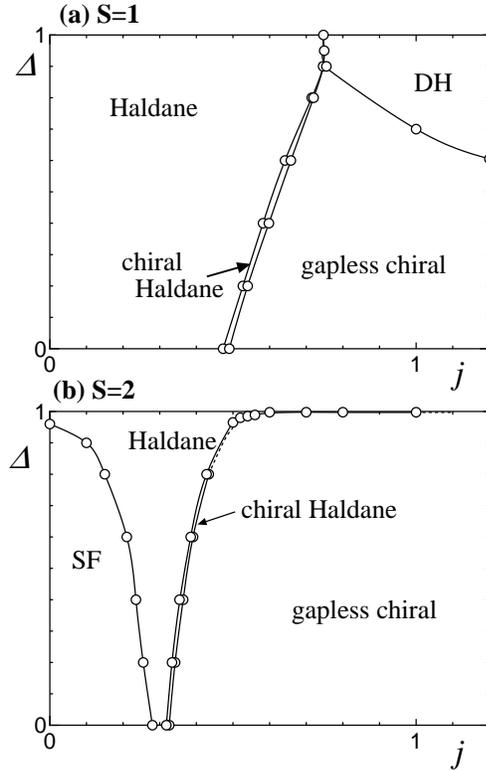}
  \caption{The ground-state phase diagram of the model (\ref{eq:HamXXZ}) 
  for (a) $S = 1$ and (b) $S = 2$ 
  in the $j \equiv J_2/J_1$ versus $\Delta$ plane.
  The circles represent the estimated transition points.
  The lines are guide for eyes.}
  \label{fig:IntS}
\end{wrapfigure}

\subsection{Integer-$S$ case}

In Fig. \ref{fig:IntS}, we show the ground-state phase diagram 
determined numerically for $S = 1$ and $2$.\cite{KKH,HKK1,HKK2}
We note that in the $S = 1$ case 
the SF phase is not present for $\Delta \ge 0$ 
but we have found it in the region of $\Delta < 0$.
This results is consistent with the bosonization study 
which suggests that for $\Delta = 0$ the SF phase appears 
in the cases of general $S$ except for $S = 1$.\cite{Leche}
Furthermore, in the $S = 2$ case the DH phase has not been found 
in the studied range of $j$, $j \le 1$, although it might exist in 
the region of larger $j$.
This issue on the appearance of the DH phase for $S \ge 2$ 
is open for future studies.

An interesting finding obtained in the DMRG studies is that 
there exist two different types of the chiral phase 
in the integer-$S$ case.
In the gapless chiral phase, the chiral LRO exists and 
the spin and string correlation functions decay algebraically
indicating gapless excitations.
In the chiral-Haldane phase, on the other hand, 
the chiral and string LROs coexist and 
the spin correlation decays exponentially indicating a finite energy gap.
The chiral-Haldane phase does exist in a very narrow but finite region 
between the Haldane and gapless chiral phases.
The appearance of the chiral phase with a finite energy gap 
has been supported later by analytic studies.\cite{Leche,Kole}

\subsection{Half-odd-integer-$S$ case}

\begin{wrapfigure}{r}{6.6cm}   
  \epsfxsize=6.6cm
  \epsfbox{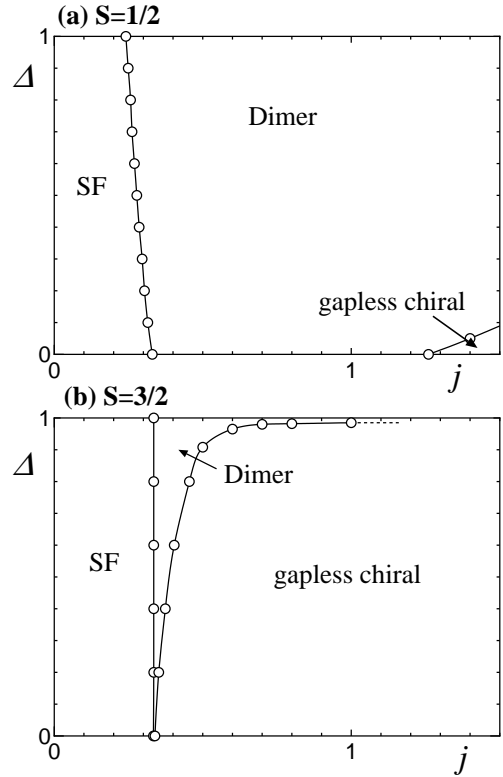}
  \caption{The ground-state phase diagram of the model (\ref{eq:HamXXZ}) 
  for (a) $S = 1/2$ and (b) $S = 3/2$ 
  in the $j \equiv J_2/J_1$ versus $\Delta$ plane.
  The circles represent the estimated transition points.
  The lines are guide for eyes.}
  \label{fig:oddS}
\end{wrapfigure}

Figure \ref{fig:oddS} shows the ground-state phase diagram 
for $S = 1/2$ and $3/2$.\cite{HKK2}
The transition line between the SF and dimer phases for $S = 1/2$ 
was determined by Nomura and Okamoto.\cite{NO}
It can be seen in the figure that the phase diagram for half-odd-integer $S$ 
includes three phases, namely, the dimer, SF, and gapless chiral phases.
We note that the appearance of the chiral phase with a finite energy gap 
between the 
dimer and gapless chiral phases 
has been suggested by the bosonization analysis,\cite{Leche} 
while such a ^^ ^^ chiral-dimer" phase has not been identified 
by the DMRG calculations.
Further studies will be necessary to solve the problem 
whether the chiral-dimer phase exists.

\subsection{$S$-dependence}

Next we discuss common features of the phase diagrams 
for general spin-size $S$.
From the numerically obtained phase diagrams 
Figs. \ref{fig:IntS} and \ref{fig:oddS},\cite{KKH,HKK1,HKK2} 
we naturally expect that the four phases, i.e., 
the Haldane, SF, gapless chiral, and chiral-Haldane phases appear 
for general integer $S$ 
(except for the $S = 1$ case where the SF phase does not appear 
for $\Delta \ge 0$ as mentioned above) 
while the three phases, i.e., the dimer, SF, and gapless chiral phases 
appear for general half-odd-integer $S$ .
The question which we wish to know here is 
how these phases move as $S$ increases.
Based on the large-$S$ approach\cite{Kole} 
and the DMRG results,
we can deduce a plausible answer to the question.
Here we only summarize the consequences.
(1) The phases with the topological order, i.e., 
the Haldane and chiral-Haldane phases for integer $S$ and 
the dimer phase for half-odd-integer $S$ become narrower as $S$ increases 
and eventually vanish in the classical limit $S \to \infty$.
(2) The region of the gapless chiral phase grows as $S$ increases and 
converges to that of the helical phase 
in the classical limit $S \to \infty$.
(3) The region of the SF phase converges to that of the N\'{e}el phase 
as $S$ increases.
We expect that in this way the phase diagrams for finite $S$ 
smoothly approach the one in the classical limit $S \to \infty$.

\section{Phase Diagram of Frustrated $S = 1$ Chain 
with Single-Ion-Type Anisotropy}

Next let us consider the frustrated quantum spin chain 
with the uniaxial single-ion-type anisotropy.
The Hamiltonian is written as 
\begin{equation}
{\cal H} = \sum_{\rho=1,2} J_\rho \sum_l 
             \left( \mib{S}_l \cdot \mib{S}_{l+\rho} \right) 
         + D \sum_l \left( S^z_l \right)^2.
\label{eq:HamD1}
\end{equation}
\begin{wrapfigure}{r}{6.6cm}   
  \epsfxsize=6.6cm
  \epsfbox{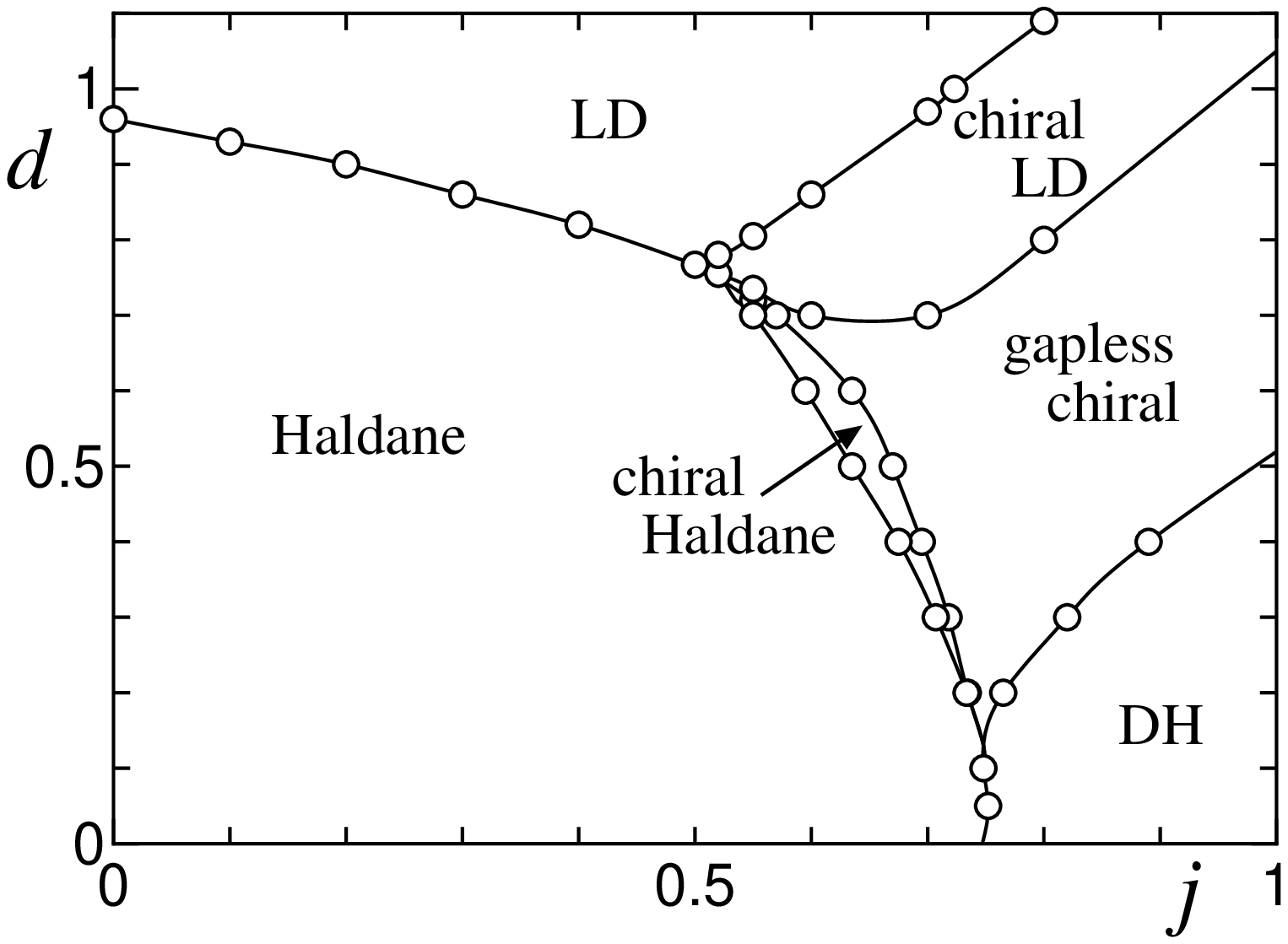}
  \caption{The ground-state phase diagram of the model (\ref{eq:HamD1}) 
  for $S = 1$ in the $j \equiv J_2/J_1$ versus $d \equiv D/J_1$ plane.
  The circles represent the estimated transition points.
  The lines are guide for eyes.}
  \label{fig:D1}
\end{wrapfigure}
Hereafter we denote $d \equiv D/J_1$ and concentrate on the case of $S = 1$.
Since the single-ion-type anisotropy is more common in real materials 
than the anisotropic exchange couplings, it is important 
to study the effect of the single-ion-type anisotropy 
for the sake of exploring experimental implications 
of the theoretical results.
For this purpose, we have determined the phase diagram of model 
(\ref{eq:HamD1}) using the DMRG method.\cite{HKK3}
The result is shown in Fig. \ref{fig:D1}.
We can see in the phase diagram that, 
in addition to the Haldane, DH, and large-D (LD)\cite{LD} phases 
which have been known, 
the chain (\ref{eq:HamD1}) exhibits three different types of 
the chiral ordered phase, i.e., the gapless chiral, 
chiral-Haldane, and chiral-LD phases.
The chiral-LD phase, which is regarded as a new type of the chiral phase 
with a finite energy gap, is characterized by the chiral LRO and 
the exponential decay of both the spin and string correlation functions.

As seen in the phase diagram Fig. \ref{fig:D1}, 
the anisotropy $d \gsim 0.15$ is large enough to realize 
the gapless chiral or chiral-Haldane phases 
if one can tune the next-nearest neighbor coupling $j$ 
within a suitable range around $j \simeq 0.75$.
These values of $d$ and $j$ are realistic so that 
there must be a good chance to find out a material in the parameter region.
Once such a material is prepared, 
the measurement of the vector chirality (\ref{eq:Ochl}) is, 
in principle, possible by using polarized neutrons.~\cite{pn1,pn2,pn3}
We hope that the results shown in the present paper stimulate 
further experimental studies to find the chiral ordered phases 
in real materials.

\section{Summary and Discussion}

In this paper, we have reviewed the recent studies 
on the ground-state properties of 
the frustrated anisotropic quantum spin chains 
(\ref{eq:HamXXZ}) and (\ref{eq:HamD1}).
From the intensive analytic and numerical studies, 
it has been shown that both of the chains exhibit the chiral ordered phases, 
where only the parity symmetry is broken spontaneously with 
preserving the time-reversal and translational symmetries.
It has also been found that there are different types of the chiral phase, 
i.e., the gapless chiral, chiral-Haldane, and chiral-LD phases.
The ground-state phase diagrams of the chains in the $j$ versus $\Delta$ 
or $j$ versus $d$ plane have been determined numerically.

To conclude this paper, we mention some issues which have been left unclear.
One of them concerns the appearance of the other types of the chiral phase.
Since the mechanism generating the chiral LRO is essentially distinct from 
the one generating the energy gap,\cite{Leche,Kole} 
there is always a possibility that a chiral phase with a finite energy gap 
appears between a gapful phase and the gapless chiral phase.
Thus, four chiral phases, i.e., the chiral-Haldane, chiral-LD, 
^^ ^^ chiral-dimer", and ^^ ^^ chiral-DH" phases are possible 
in the phase diagram of the chains (\ref{eq:HamXXZ}) and (\ref{eq:HamD1}).
Among them, the ^^ ^^ chiral-dimer" and ^^ ^^ chiral-DH" phases have not been 
identified in the DMRG calculations.
We must say, however, that it is difficult to exclude the possibility 
that a phase exists in a too narrow region to be detected 
using a numerical method.
Further studies will be required to clarify the fate of the phases.

Another important problem might be the finite-temperature properties 
of the chiral phases.
Some knowledge on excitations have been suggested 
by analytic studies.\cite{Ner,Leche,Kole}
For example, in the gapless chiral phase 
the effective Hamiltonian for the low-energy gapless and gapful modes 
are represented as the Tomonaga-Luttinger liquid 
and the sine-Gordon model, respectively.
Nevertheless, since these analytic results are based on approximations 
inevitable in the approaches, a check by other methods such 
as numerical ones is needed.
Furthermore, quantitative studies, which are important for 
experimental implications, are still lacking.
More efforts including numerical ones are desirable to tackle 
the problem.

\section*{Acknowledgements}
The authors are very grateful to T. Tonegawa for fruitful discussions.
Numerical calculations were carried out in part 
at Yukawa Institute Computer Facility, Kyoto University.
T.H. was supported by a Grant-in-Aid for Encouragement 
of Young Scientists from Ministry of Education, Science and Culture of Japan.

\end{document}